# Graphene/ TiS$_3$ heterojunction for selective polar vapor sensing at room temperature

Nassim Rafiefard[a], Azam Iraji zad*[a,b], Ali Esfandiar[b], Pezhman Sasanpour[c], Somayeh Fardindoost [b] Yichao Zou[d], Sarah.J.Haigh[d], Seyed Hossein Hosseini [bf]

[a] Institute for Nanoscience and Nanotechnology, Sharif University of Technology, Tehran 14588-89694, Iran
[b] Department of Physics, Sharif University of Technology, Tehran 11155-9161, Iran.
[c] Department of Medical Physics and Biomedical Engineering, School of Medicine, Shahid Beheshti University of Medical Sciences, Tehran, Iran
[d] School of Materials, The University of Manchester, Manchester M13 9PL, UK
[f] Department of Physics, Iran University of Science and Technology, Tehran, 13114-16846, Iran
Corresponding author: Iraji@sharif.edu

**Abstract**

In this work, the room temperature polar vapor sensing behavior of two dimentional (2D) heterojunction Graphene-TiS$_3$ materials and TiS$_3$ nanoribbons is investigated. TiS$_3$ nanoribbons were synthesized via chemical vapor transport (CVT) and their structure was investigated by scanning electron microscopy (SEM), high resolution transmission electron microscopy (HRTEM), energy dispersive X- ray spectroscopy (EDS), X-ray diffraction (XRD), Raman spectroscopy and Fourier transform infrared spectroscopy (FT-IR) analysis. The gas sensing performance of the TiS$_3$ nanoribbons was assessed through the observed changes in their electronic behavior. Sensing devices were fabricated with gold contacts and with lithographically patterned graphene (Gr) electrodes in a 2D heterojunction Gr-TiS$_3$-Gr architecture.. It is observed that the gold contacted TiS$_3$ device has a rather linear I-V behavior while the Gr-TiS$_3$-Gr heterojunction forms a contact with a higher Schottky barrier (250 meV). I-V responses of the sensors were recorded at room temperature with a relative humidity of 55% and for different ethanol vapor concentrations (varying from 2 to 20 ppm). I-V plots indicated an increase in the resistance of Gr-TiS$_3$-Gr by the adsorption of water and ethanol molecules with relatively high sensing response (~3353% at 2 ppm). Our results reveal that selective and stable responses to a low concentration of ethanol vapor (2 ppm) can be achieved at room temperature with transient response and recovery times of around 6 s and 40 s, respectively. Our proposed design demonstrate a new approach for selective molecular recognition using polar interactions between analyte vapors and heterojunctions of 2D-materials.

**Introduction**

The number of successfully exfoliated 2D semiconductor materials is rapidly expanding and attracting considerable attention as a result of their unique physical and chemical properties that vary as a function of physical and chemical characteristics [1-4]. Numerous studies have revealed unique electrical [5-9], optical [8,10-13], thermal [14] and mechanical [15] properties that do not exist in the bulk counterparts. Among the broad family of these layered materials, transition metal dichalcogenides (TMDCs) have advantages for gas sensing due to their high surface to volume ratio and the existence of functional groups on their edges and surfaces [16-20]. For instance, single and few-layer $MoS_2$ sheets have been demonstrated to be sensitive detectors for NO, $NO_2$, $NH_3$, organic and triethylamine gases, due to the n or p-type doping induced by the adsorbed gas molecules [21-24].

Less widely studied 2D semiconductors are the family of transition metal trichalcogenides (TMTCs) which includes $ZrS_3$, $ZrSe_3$, and $TiS_3$ [8,25]. The $TiS_3$ has chalcogen trimers at flake edges, which are potentially potent sites for gas sensing applications [26]. In our previous study, we predicted the gas sensing and selectivity of $TiS_3$ nanoflakes at room temperature through density functional theory (DFT) calculations [27]. We computationally obtained a profile for the adsorption intensity of analyte including polar and non-polar molecules on different surface facets of $TiS_3$. The simulations showed that the adsorption of hydroxyl (OH) groups on S-vacancy sites on the (001) surface is energetically favorable and generally stronger than other facets due to the bipolar interactions between the molecule and the surface.

Here, using graphene (Gr) electrodes as 2D-contacts to a $TiS_3$ flake grown by the chemical vapor transport (CVT), we experimentally demonstrate enhancement in ethanol sensing at room temperature. I-V measurements are used to understand the mechanism of gas interaction with the sensing material through changes in electronic behavior. We observe that the Gr electrodes in contact with the active $TiS_3$ channel play an important role in determining the responsivity of our fabricated gas sensor [44]. The insights achieved from this study may provide new benchmark for the use of hybrid 2D-materials as gas sensors.

**Results and discussion**
**Growth and characterization of TiS$_3$ and graphene crystals**

Long ribbons of TiS$_3$ have been grown by CVT via direct reaction of metallic titanium and elemental sulfur in a reaction ampule [28]. In this process, the reaction takes place between sulfur powder (AnalaR NORMAPUR, 99.5 %) and a 0.25 mm thick Ti foil (Alfa Aesar, 99.5 %), with a molar S:Ti ratio of 3:1, inside a vacuum sealed quartz ampule (p~80 mTorr) at 450 °C for 72 h, (Fig. 1. a). As can be seen, sulfur is visible as the yellow powder inside the ampule in the vicinity of the Ti foil. Fig. 1b provides optical images of the ampule at the end of the process when it is cooled to room temperature. The pieces of Ti foil are covered by a dense forest of TiS$_3$ ribbons as illustrated by the scanning electron microscope (SEM) images in Fig. 1c and d. The un-reacted dark brown sulfur is condensed on the walls of the ampule shown by the red arrow in Fig. 1b. The TiS$_3$ ribbons are typically over 100 μm in length, ~5 μm wide and hundreds of nanometers in thickness. Fig. 1e shows the high angle annular dark field (HAADF) scanning transmission electron microscope (STEM) images and corresponding energy dispersive spectroscopy (EDS) elemental maps of the TiS$_3$ ribbon (Fig. 1 f-h). The STEM - EDS imaging was performed using a Talos X-FEG STEM with a operating at an accelerating voltage of 200 kV and a total EDS collection solid angle of 0.9 srad. HR-TEM imaging and electron diffraction was performed with an FEI F30 STEM operated at 300 kV. The composite elemental map for Ti and S in Fig. 1h suggests that there is a thin S-deficient layer at the edges which is also enriched in oxygen. Quantification of the sum EDS spectrum from the center of the ribbon indicates an average atomic fraction of Ti to S equal to 3, consistent with the stoichiometry of TiS$_3$ (data overlaid in Fig.1h).

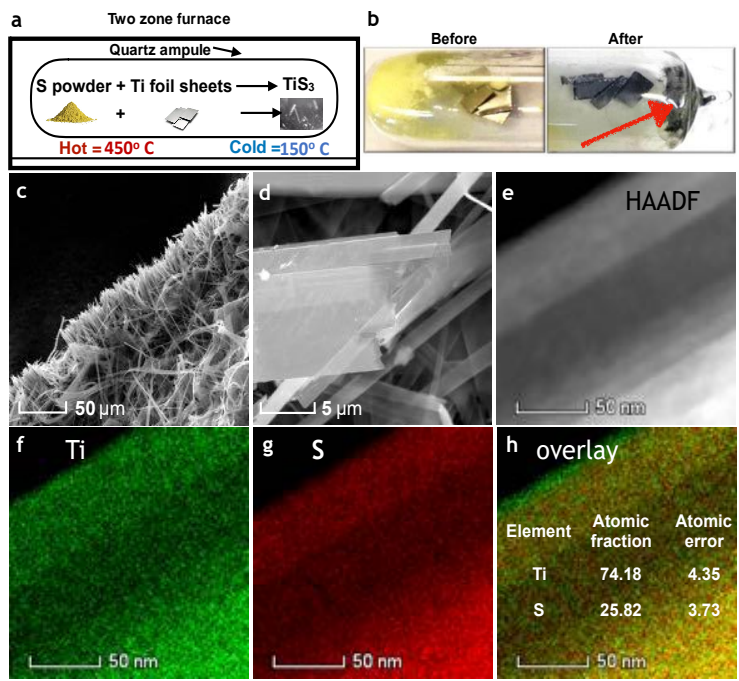

**Fig 1. Synthesis and characterization of TiS₃ ribbons**. a) Schematic and b) photograph of the reaction ampule containing yellow sulphur powder and Ti foil before synthesis and with TiS₃ ribbons covering the remaining Ti foil and condensed S after the synthesis. (c and d) SEM images of TiS₃ ribbons densely grown on the Ti foil at different magnifications. Ribbons are approximately 100 µm in length, a maximum of 5 µm in width and a hundred nanometers in thickness. (e-h) STEM HAADF image and corresponding EDS maps of a TiS₃ ribbon. The quantified mean elemental ratio for the mapped region is shown in the overlay image

Fig. 2 shows the high-resolution transmission electron microscope (HR-TEM) images and corresponding selective-area electron diffraction (SAED) patterns obtained from the TiS₃ ribbons. The image in Fig. 2a-c shows a ribbon lying flat on the carbon support in which the lattice spacings and directions match the <010> and <100> TiS₃ lattice directions, demonstrating that the ribbon is viewed along [001] (Table.1). Fig. 2d-f shows a ribbon positioned lying on its edge; here the lattice spacings and directions match <001> and <010>, demonstrating the edges of the TiS₃ ribbons are oriented parallel to the (100) crystal planes [27].

Table 1. Lattice constants of bulk TiS3 (calculated and experimental results).

|  | a (Å) | b (Å) | c (Å) |
|---|---|---|---|
| expt.[1] | 4.9 | 3.37 | 8.7 |
| calc.[27] | 4.9 | 3.38 | 8.8 |
| This work | 4.9 | 3.4 | 8.7 |

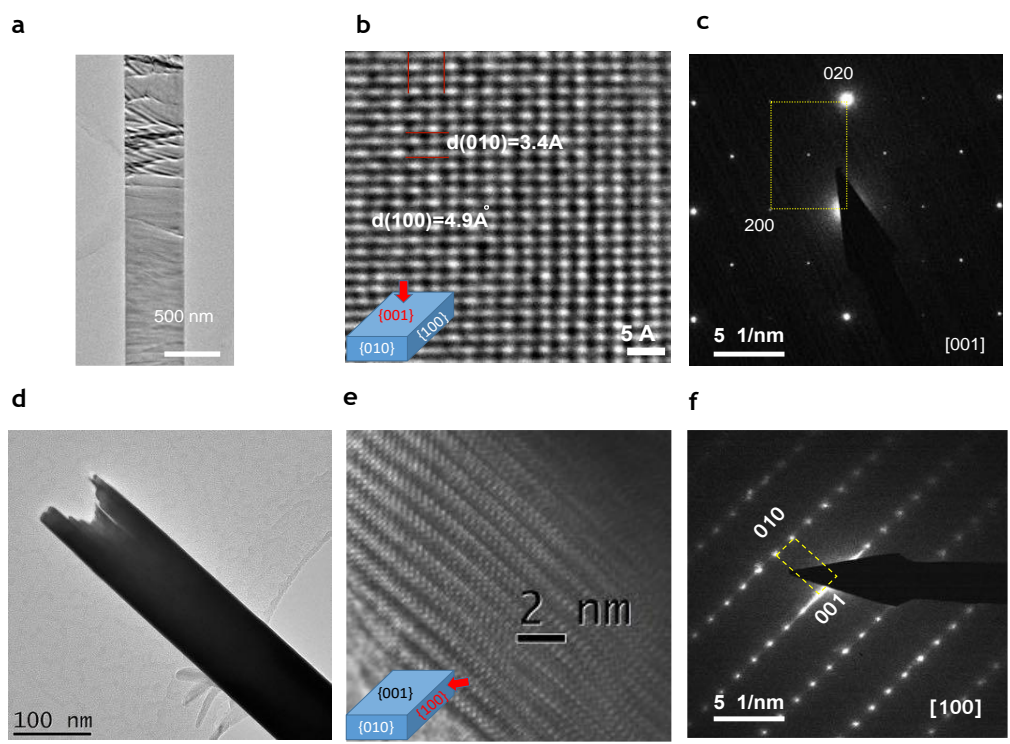

**Fig. 2 Transmission electron microscopy characterization of TiS₃ nanoribbons**. a) Bright-field (BF) TEM image taken from a section of a typical TiS₃ nanobelt; (b) Corresponding selective-area electron diffraction (SAED) pattern from the thin edge of the nanobelt, taken along [001] zone-axis. Since the nanobelt is naturally lying flat on the grid, the electron beam direction should be parallel to its surface normal. Combining (a) and (b), the axial growth direction of the nanobelt is determined as [010]. (c) Corresponding high-resolution (HR) TEM Image. d) BF TEM image taken from [100] section. e) Corresponding selective-area electron diffraction (SAED) pattern from the thin edge of the nanobelt, taken along [100] zone-axis. f) Corresponding high-resolution (HR) TEM image.

## Fabrication of TiS₃ and graphene devices

The TiS₃ nanoribbons were fabricated using traditional metallic contacts (named as Au-TiS₃) and with graphene as an intermediate contact to the TiS₃ (referred to as Au-Gr-TiS₃). Both structures are shown schematically in Fig 3 a. The fabrication procedure briefly was as following: The TiS₃ ribbons were exfoliated and transferred onto a SiO₂ (300 nm)/ p++Si substrate by mechanical exfoliation from a bunch of TiS₃ ribbons [29]. For the Au-TiS₃ device, Au/Cr electrodes were evaporated (3 nm of Cr followed by 50 nm of Au) contacting either end of the ribbon as shown in Fig. 3a (upper panel). To fabricate the Au-Gr-TiS₃ device, we used graphene sheets grown via chemical vapor deposition (CVD) on copper foils as reported previously [30]. After removal of the copper substrate, the Gr flake was transferred onto a SiO$_2$(300nm)/p+Si wafer. A 20 µm wide stripe was then etched in the CVD-Gr using standard photolithography processes and oxygen plasma etching. The etched Gr sheet was then transferred to a TiS$_3$ ribbon and aligned such that the ribbon bisected the gap in the Gr sheet. To electrically contact the graphene and complete the device, two Au (50 nm) / Cr (3 nm) electrodes were then evaporated on to the Gr on each side of the TiS$_3$ ribbon to complete the Au-Gr-TiS$_3$ devices as shown in Fig. 3a, (lower panel). Fig. 3b-c show the optical images of the final devices. To check the quality of prepared devices, we have performed confocal Raman analysis (532 nm laser wavelength, Xplora™ Plus- HORIBA), as presented in fig 3d-f. In Fig 3d the sharp peak of 2D and the high ratio of the 2D/G bands demonstrates that the electrodes are composed of high quality single layer graphene.

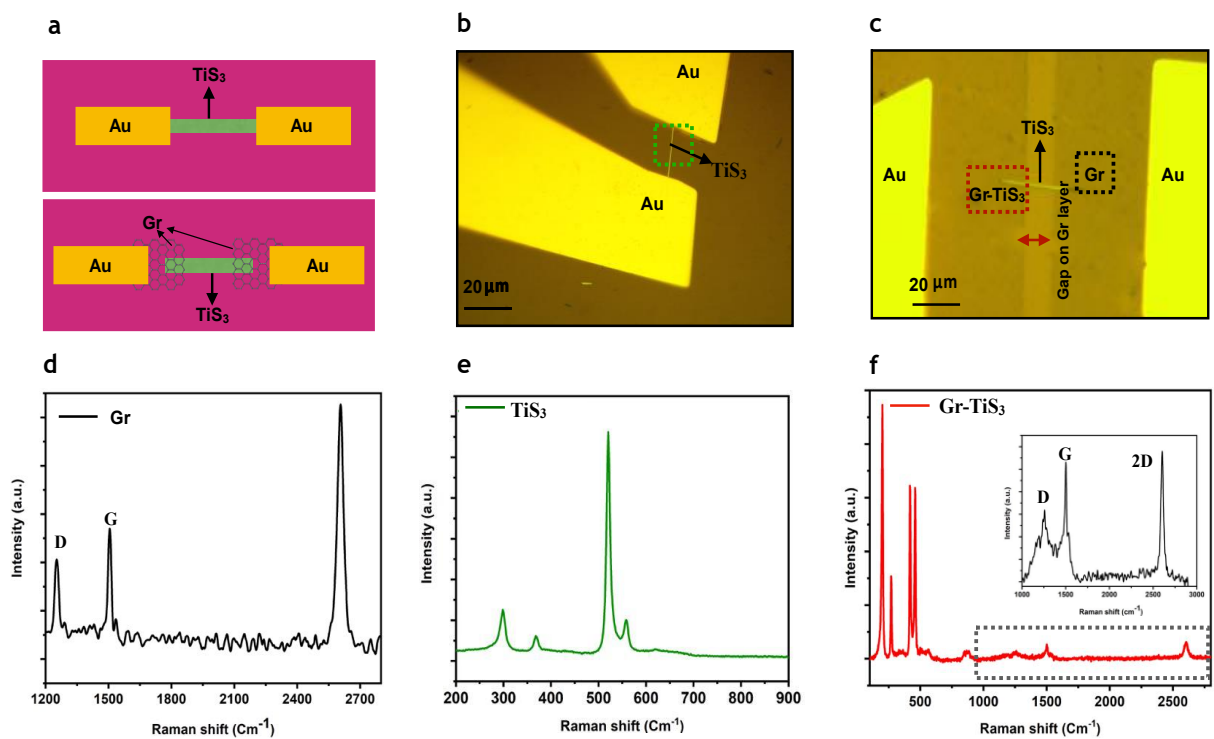

**Fig 3. Sensor device fabrication and characterization.** a) Schematic of transferred TiS$_3$ and Gr on Si/SiO$_2$ substrate. Optical images of b) Au-TiS3-Au c) Au-Gr-TiS3-Gr-Au devices. Raman spectrum from d) Gr e) TiS$_3$ f) Gr-TiS$_3$ heterojunction regions of the Au-Gr-TiS$_3$-Gr-Au device.

The Raman spectra collected from the TiS$_3$ nanoribbon shows 5 prominent peaks (Fig. 3e). The 4 peaks at 177 cm$^{-1}$, 300 cm$^{-1}$, 371 cm$^{-1}$, and 559 cm$^{-1}$ correspond to A$_g$-type Raman modes of the

TiS$_3$ crystal and are in good agreement with reported modes for TiS$_3$ flakes [31] while the peak at ~520 cm$^{-1}$ comes from the underlying silicon substrate.

The Raman spectrum of Gr-TiS$_3$ in Fig. 3f reveals presence of both Gr and TiS$_3$ characteristic peaks indicating Gr-TiS$_3$ heterojunction region with no considerable changes.

**Electrical measurement**

The electrical characterization of the devices was carried out in ambient conditions using Keithley Series 6400 Pico-ammeters from -1 V to 1 V in the step of 0.1 V and delay of 0.05 s. Fig.4a shows the current–voltage characteristics (I$_{ds}$-V$_{ds}$) of Au-Gr-TiS$_3$ and Au-TiS$_3$ devices. The Au in contact with TiS$_3$ has a rather linear I-V response with a resistance of 390 kΩ, while in the case of the Au-Gr-TiS$_3$ device, the device has a nonlinear response with a higher resistance of 2500 kΩ. As depicted in the schematic diagram of fig.4b, the position of the TiS$_3$ conduction band is lower than the Fermi level of Au, so the electrons migrate freely from Au to TiS$_3$ nanoribbons. On the other hand, as reported by Liu et al., the work function of the Gr-TiS$_3$ heterojunction is about 4.9 eV [32], higher than the Fermi level of the Au metal which results in a Schottky barrier.

Fig. 4b shows estimated band diagrams for the two devices before and after contacting. We estimate the Schottky barrier height from the I-V characteristics to be 250 meV in the case of Au-Gr-TiS$_3$ which correlates with the theoretical value previously predicted [32]. The contact potential difference, $\Delta V$, between Gr and TiS$_3$ (2.34 eV) is greater than that between Au and TiS$_3$ (2.15eV), which is another reason for the observed difference in the Schottky barrier height. In the case of Au-TiS$_3$, the interaction between the Au and the nanoribbon is expected to cause metal induced defect energy levels in the TiS$_3$ semiconductor that induce Fermi level pinning (FLP) at the contact. In contrast in Au-Gr-TiS$_3$ the graphene and semiconductor components are expected to have a weak van der Waals interaction, producing minimal metal induced defect energy levels leading to negligible FLP. The absence of FLP and the greater Schottky barrier height of Au-Gr-TiS$_3$ compared to Au-TiS$_3$ devices is likely to be more which is desirable in sensing process [32].

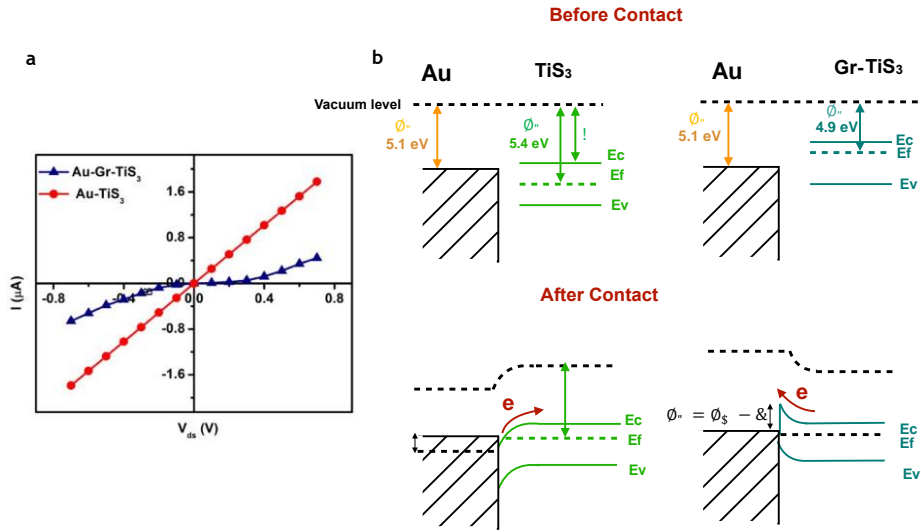

**Fig. 4. Electrical measurements.** a) I-V curves of the Au-Gr-TiS$_3$ and Au-TiS$_3$ devices b) Schematic band diagram for the Au-TiS$_3$ and Au-Gr-TiS$_3$ devices before and after contacting.

**Ethanol vapor sensing properties of Gr-TiS$_3$ heterojunction**

In order to investigate the room temperature polar vapor sensing behavior of the Au-Gr-TiS$_3$ and Au-TiS$_3$ devices, I-V characteristics were recorded in the presence of different vapor concentrations. The fabricated sensors on portable chip carrier were placed on top of a sensing bowl connected to an I-V multimeter which is logged to a computer as depicted schematically in Fig.5a. A range of ethanol vapor concentrations were made by mixing different volume ratios of absolute ethanol (from Merck Ltd., 99.9%) with DI-water (18 MOhm-cm) followed by flowing dry air into the mixture at room temperature. We used Henry's law to calculate the concentration of the ethanol vapor in ppm [33]. The sensor response was defined as the relative change in the current ($I_{ethanol}$–$I_{air}$)/$I_{air}$ at a constant voltage, where $I_{ethanol}$ and $I_{air}$ denoting the current in the presence of ethanol vapor and in air, with relative humidity (RH) of 55% and 30% respectively. Fig. 5b shows the I-V characteristic of the Au-TiS$_3$ nanoribbon when exposed to 2-20 ppm ethanol vapor at a RH of 55%. In our set-up for gas sensing, there are three factors that could affect the sensors' I-V response: the rate of dry air flow; the bowl humidity; and the ethanol vapor pressure. The sensing response was assessed when exposed to different rates of dry air flow and the I-V response did not show any change. When the Au-TiS$_3$ device was tested for only a change in the humidity of air from 35% to 55% RH the resistance decreased from 390 kΩ to 350 kΩ (at V=0.6V) (as shown in Fig. 5b). Then the sensor was exposed to the different concentrations of ethanol vapor from 2 to 20 ppm all at a constant relative humidity (55% RH). As can be seen in Fig. 5b, the presence of ethanol vapor decreases the resistance more than simply increasing the relative humidity. The sensor responses to air and 2 ppm ethanol are 16% and 24%, respectively. In the case of the Au-TiS$_3$ device, the differences between the ethanol vapor concentrations, from 2 ppm to 20 ppm, is small and unstable which prevents sensing of the concentration of ethanol vapor.

In contrast, when exposing Au-Gr-TiS$_3$ device to ethanol vapor at RH=55%, produced an impressive increase in the sensor's response (Fig. 5c). The sensor's responses to air and 2 ppm ethanol are 517% and 3353%, respectively. In the case of 20 ppm ethanol at RH=55% the sensor's response is 3400%. Fig. 5e, compares the sensing response to 6ppm ethanol of the two TiS$_3$ devices (Au-Gr-TiS$_3$ and Au-TiS$_3$) with a pure graphene control. This demonstrates the much greater sensitivity of the Au-Gr-TiS$_3$ heterojunction compared to the Au-TiS$_3$, while intrinsic graphene alone has no sensing response for ethanol which is in agreement with previous works [34].

Next, we tested the dynamics of the sensor for a 6ppm ethanol vapor. As seen in Fig. 5d, the Au-Gr-TiS$_3$ sensor shows a relatively fast response when exposed to the vapor as well as good recovery time when the vapor is removed (response and recovery times were 6 s and 40 s respectively).

The selectivity of the Au-Gr-TiS$_3$ device to humidity, ethanol, methanol, acetone, H$_2$, CH$_4$, and CO was investigated as shown in Fig. 5f. We found that there is no significant sensing response even to high concentrations (>1000 ppm) of small molecule gas species such as H$_2$, CH$_4$, CO.

Ethanol showed a higher response than acetone, which is explained by the stronger absorption behavior resulting from the bipolar nature of the hydroxyl (OH) group, which interacts with the surface via two distinct charge transfer modes [35]. Our results confirm that the polarity and the presence of OH group in analytes is key factor for the selective polar vapor sensing of TiS$_3$ at room temperature.

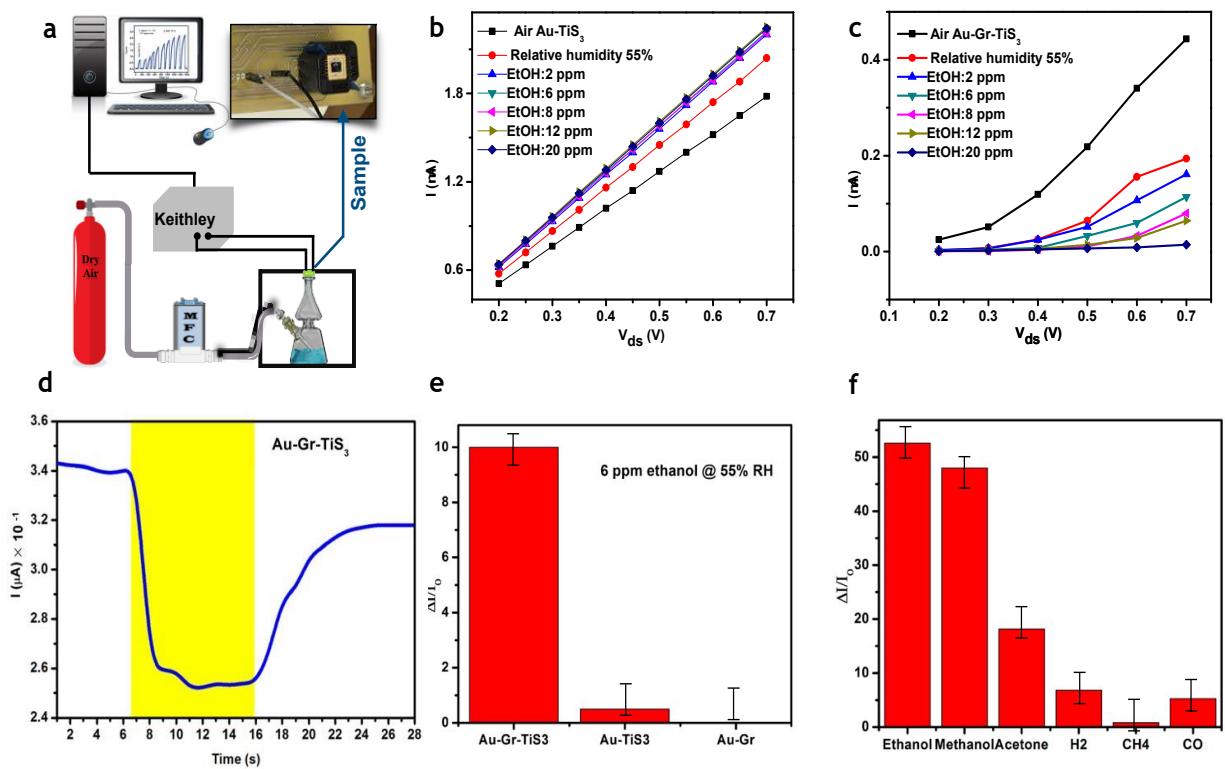

**Fig. 5. Vapor sensing data**. a) Schematic diagram of the gas sensor experimental set-up b) I-V characteristic of the Au-TiS$_3$ device when exposed to 2-20 ppm ethanol vapor for a relative humidity (RH) of 55 % at room temperature. c) I-V characteristic of the Au-Gr-TiS$_3$ heterojunction device when exposed to 2-20 ppm ethanol vapor for a relative humidity (RH) of 55 % at room temperature. d) Dynamic response of Au-Gr-TiS$_3$ to 6 ppm ethanol. The yellow part refers to the period when the sample is exposed to ethanol. d) Comparing the responses to 2 ppm ethanol for Au-TiS$_3$, Au-Gr-TiS$_3$ and Au-Gr devices e) the Au-Gr-TiS$_3$ sensor response to ethanol, methanol, acetone, H$_2$, CO and CH$_4$ indicating selective response to the target gases only.

The rate of absorption of a gas molecule on the surface of TiS$_3$ is affected by several parameters including the presence of sulfur (S) vacancies, defect states and unstable bonds [36]. Our HR-TEM data demonstrated that the largest surfaces of the TiS3 nanoribbons are oriented along

(001). Our previous computational studies suggested that in an anaerobic environment, $C_2H_5OH$ and $H_2O$ molecules react with Ti and S vacancies on $TiS_3$ (001) surfaces by direct adsorption of the bipolar hydroxyl (OH) group through binding of an oxygen atom to both Ti and S atoms [27]. We demonstrated that the OH group of the polar molecule binds more strongly to a Ti than a S atom.

To investigate the behavior of the OH stretching band in water and ethanol molecules adsorbed on the surface of $TiS_3$, we used FTIR spectroscopy in the wavenumber region 500–4000 cm$^{-1}$. Fig. 6a compares the IR absorption spectra of our $TiS_3$ nanoribbons exposed to water and ethanol to the case when the sample is just in air in order to evaluate the intensity ratio of the OH stretching bands adsorbed on the $TiS_3$ surface. In presence water molecules, the OH groups show an extremely wide and non-homogeneously broadened band stretching line in the region 3000– 3800 cm$^{-1}$. Upon exposing to the ethanol, this wide region is weakened which has been investigated in detail by Burikov et al., who demonstrated this change is caused by the formation of chain aggregates from ethanol/water molecules [37]. Other adsorption peaks for wavenumbers of 1064 cm$^{-1}$, 2916 cm$^{-1}$, correspond to C-O, C-H stretching bands in the ethanol molecule. This confirms that the $TiS_3$ is responding to the ethanol, with the adsorbed ethanol and water molecules interacting with the Ti and S vacancies ($V_{Ti/S}$) as it can be expressed as following equations [22]:

$C_2H_5OH$ (gas) + $V_{Ti/S}$ 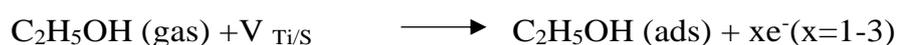 $C_2H_5OH$ (ads) + $xe^-$ (x=1-3)

$H_2O$ + $V_{Ti/S}$ 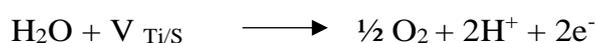 ½ $O_2$ + $2H^+$ + $2e^-$

The reaction mechanism is illustrated schematically in Fig. 6a, with the released electrons inside the $TiS_3$ transferring to the Au and resulting in a decrease in resistance, which is in agreement with the measured I-V characteristic of the Au-$TiS_3$ sample (Fig. 5b).

To understand the improved sensing performance of the Au-Gr-$TiS_3$ heterojunction, a charge carrier depletion region or energy barrier created at the $TiS_3$-Gr interface must be considered. Liu et al. studied the carrier transport at the interface of $TiS_3$ with a Gr monolayer, considering the carrier injection efficiency by analyzing the tunnel barrier, band bending, and the Schottky barrier height [32]. They found that among these the Schottky barrier at the $TiS_3$-Gr junction is the most important factor for determining the contact resistance when the $TiS_3$ is exposed to the polar vapor molecules. We measured the Schottky barrier height as about 250 meV for Au-Gr-$TiS_3$ which produces a small amount of band bending in the charges flow path when traveling from the Gr-$TiS_3$ region to the Au contact (Fig. 4b). According to equations 1 and 2, when exposing the Au-Gr-$TiS_3$ to polar vapors, the induced carriers in the $TiS_3$ experience a repulsive force from each other and tend to move from a high concentration region ($TiS_3$) to a lower concentration region (Gr) before joining the positive ion. This results in the widening of the depletion region at the $TiS_3$ and Gr interface and increases the height of the Schottky barrier followed by an increase in the resistance of the Au-Gr-$TiS_3$ device. The Au-$TiS_3$ device showed only a small resistance variation on exposure to the ethanol molecules due to the absence of any barrier in the flow path of the charge carriers. So, the charge carrier depletion or an energy barrier created at the $TiS_3$-Gr junction induces increased band bending, with a higher Schottky barrier, resulting in enhanced sensing for polar molecules at room temperature for graphene contacted $TiS_3$ compared to gold contacted $TiS_3$. The role of Gr as electrode for 2D materials not only improves the contact resistance in some cases, but also we showed that it can creates an extra sensitive junction for polar vapor detection and discrimination.

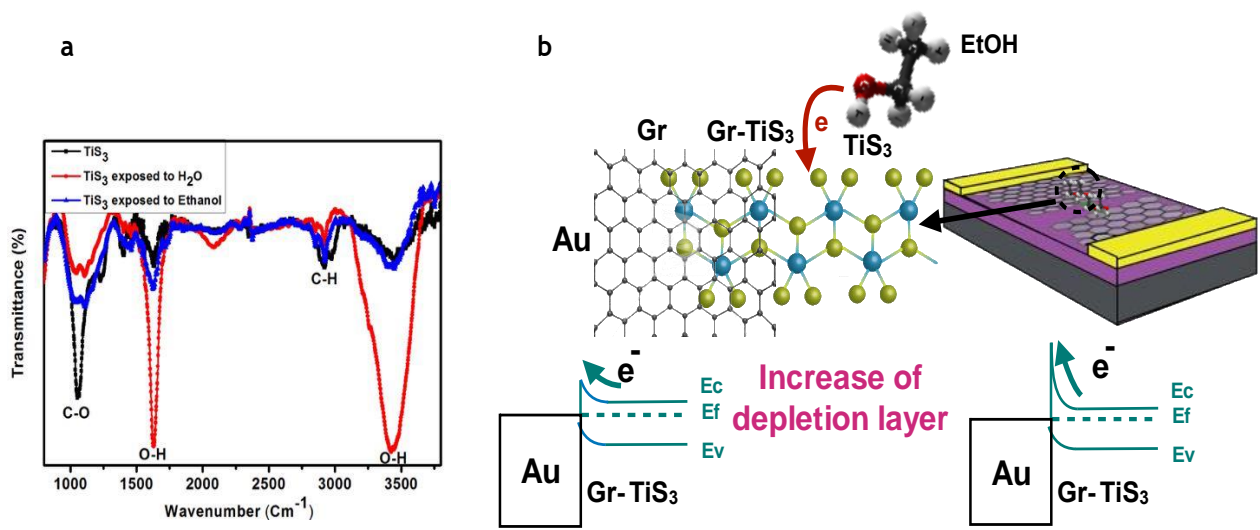

**Fig. 6. Ethanol sensing mechanism.** a) The FT-IR absorption spectra of adsorbed water and ethanol on the $TiS_3$ nanoflakes within the region of the CH and OH stretching bands. b) Schematic of the gas adsorption mechanism and band bending as charge carriers move from the $TiS_3$-Gr region to the Au contact before and after gas exposure.

## Conclusion

In summary, we have demonstrated the potential of a Gr-$TiS_3$ heterojunction device for ethanol and polar vapor sensing with a fast, selective and reversible response at room temperature. The sensing response was as high as 3353% when the graphene contacted $TiS_3$ was exposed to 2 ppm ethanol compared to 24% for the same $TiS_3$ nanoribbons with Au contacts. The sensing mechanism can be explained based on weak chemisorption of ethanol assisted with water molecules on the $TiS_3$ nanoribbons (001) surface, followed by electron transport. The superior performance of our heterojunction is attributable to the Schottky barrier formed between the $TiS_3$ and the Gr flake. Our results indicate that the proposed hybrid structure is more promising approach for detection of volatile organic molecules which overcomes the limitations of Gr and $TiS_3$ when used individually.

## Competing interests

The authors declare no competing financial interests.

## Acknowledgements

A.I. thanks the National Elites Foundation (of Iran) for the financial support of his research. A.E. would like to thank the Iran National Science Foundation (INSF, Grant No. 96011388). S.J.H. thanks the Engineering and Physical Sciences (U.K.) (Grants EP/M010619/1, EP/P009050/1) and the European Research Council (ERC) under the European Union's Horizon 2020 research and innovation programme (Grant ERC-2016-STG-EvoluTEM-715502 and the ERC Synergy project). We thank Diamond Light Source for access and support in use of the electron Physical Science Imaging Centre that contributed to the results presented here.